\documentclass[mathleft]{an}
\usepackage{graphicx}
\usepackage{times}
\newcommand{\Msun}{M$_{\odot}$}
\newcommand{\Mjup}{M$_{JUP}$}

\begin{document}

\Pagespan{789}{}
\Yearpublication{2006}%
\Yearsubmission{2005}%
\Month{11}%
\Volume{999}%
\Issue{88}%

\title{Jets from Young Stars and Brown Dwarfs \thanks{}}

\author{E.T. Whelan\inst{1}\fnmsep\thanks{\email{emma.whelan@astro.uni-tuebingen.de}}
}
\titlerunning{Jets from Young Stars and Brown Dwarfs}
\authorrunning{E.T. Whelan}
\institute{Institut f{\"u}r Astronomie und Astrophysik, Kepler Center for Astro and Particle Physics, Eberhard Karls Universit{\"a}t,  Sand 1, 72076 T{\"u}bingen, Germany}

\received{}
\accepted{}
\publonline{later}


\abstract{The protostellar outflow mechanism operates for a significant fraction of the pre-main sequence phase of a solar mass star and is thought to have a key role in star and perhaps even planet formation. This energetic mechanism manifests itself in several different forms and on many scales. Thus outflow activity can be probed in numerous different regimes from radio to X-ray wavelengths. Recent discoveries have shown that it is not only solar mass stars that launch outflows during their formation but also the sub-stellar brown dwarfs. In this article what is currently known about jets from young stars is summarised, including an outline of why it is important to study jets. The second part of this article is dedicated to jets from young brown dwarfs. While only a small number of brown dwarf outflows have been investigated to date, interesting properties have been observed. Here observations of brown dwarf outflows are described and what is currently known of their properties compared to low mass protostellar outflows. }

\maketitle

\section{Jets from Young Stars}
The outflow phenomenon is ubiquitous in star forming regions and is observed from radio to X-ray wavelengths (Reipurth $\&$ Bally 2001). As a result, collimated outflows or {\em jets} have been the subject of some of the most stunning images of star formation (Figure 1). It is more than 25 years since protostellar jets were first discovered and they are now known to be central to the formation of both stellar and sub-stellar objects with the most massive jet driver observed having a mass of  $\sim$ 30~\Msun\ (Zhang et al. 2013) and the least massive a mass of only 24~\Mjup\ (Whelan et al. 2012)\footnote{New observations have hinted that outflow activity could extend to the planetary mass boundary but further observations are needed to confirm this (Joergens 2013)}. See Ray (2007) for a discussion of the first observations of protostellar jets. Jets are also detected in all pre-main sequence evolutionary stages and thus can persist for millions of years (Cabrit et al. 2011). Furthermore, the scales on which they are observed range from 10's of AU (Agra-Amboage et al. 2011) to several parsecs (McGroarty, Ray $\&$ Bally 2004). While jets have been detected from brown dwarfs (BDs, 13~\Mjup $\leq$ M$_{BD}$ $\leq$ 75~\Mjup) and massive stars (M$_{\star}$ $\geq$ 10~\Msun) much of what is understood about protostellar jets comes from the study of low mass (0.2~\Msun $\leq$ M$_{\star}$ $\leq$ 2~\Msun) young stellar objects (YSOs).

A low mass star forms via the core collapse mechanism, where a core of molecular material becomes gravitationally unstable and collapses under its own weight. The protostar grows at the centre of the core as densities and temperatures increase. After about $\sim$ 10,000 years a core-halo structure results where a disk has formed around the protostar and the star-disk system is surrounded by a cloud of material. The protostar forms through accretion of the ambient medium via the disk and the magnetospheric accretion paradigm, where accretion onto the star proceeds along magnetic field lines linking the star and disk, best describes accretion in protostars (Bouvier et al. 2007). Early in the lifetime of the protostar an outflow is launched. The star and disk are rotating and it is the rotation of this system combined with the infall of material onto the disk that results in ejection. The outflowing material is collimated into a jet by the magnetic field of the system. It is widely argued that the outflow mechanism is responsible for the removal of angular momentum from the star-disk system, thus allowing accretion to proceed and the star to form. 

\begin{figure}
\centering
\includegraphics[width=65mm, trim= 0cm 0cm 0cm 0cm]{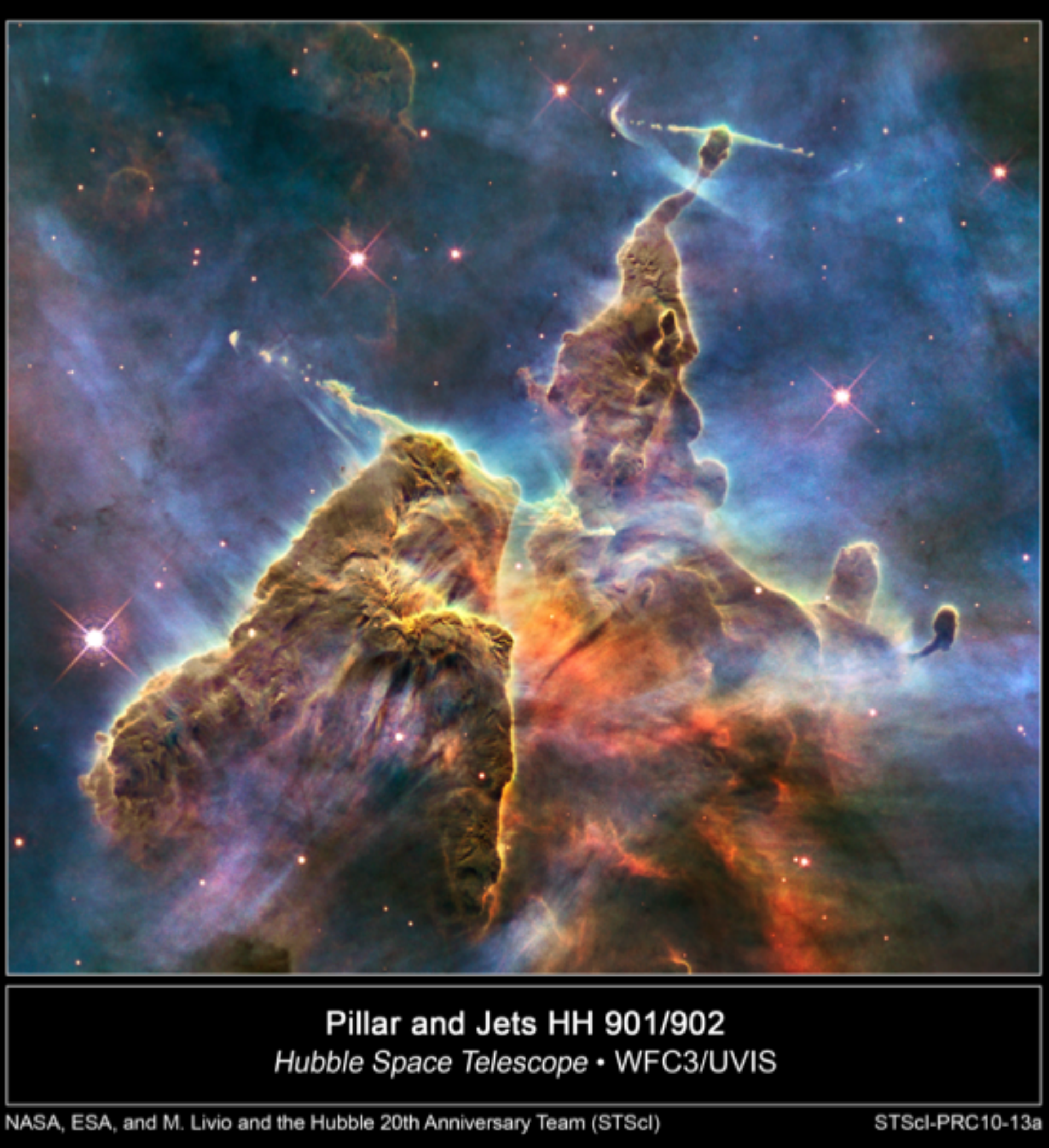}
\caption{Protostellar jets in the Carina star forming region. Note the characteristic collimation of the jets, the bow shocks and the knotty structure.}
\label{label1}
\end{figure}

The jet launching scenario detailed above is a much simplified outline of magneto-centrifugal jet launching. Magneto-centrifugal launching is accepted as the likely mechanism for the generation of protostellar jets yet the precise model is still debated with both disk winds and X-winds being strongly considered (Ferreira, Cabrit $\&$ Dougados 2006). What is known is that outflow and accretion activity are strongly connected (Cabrit et al. 2010) and it is estimated that in low mass YSOs $\sim$ 1-10$\%$ of the infalling material is ejected.  The HH~212 jet shown in Figure 2 is an archetypal protostellar jet and while such perfect jets are not often observed, protostellar jets will exhibit all of the main features of the HH~212 jet. That is the bipolarity, high degree of collimation, bow shaped shocks, and episodicity. The episodicity refers to the fact that the jet is made up of a series of shocks or knots and terminates in a bow shock. It is postulated that the each knot corresponds to a separate ejection and thus accretion event and that variable accretion is responsible for this property.

The youngest jet driving sources, e.g. Class 0, I low mass protostars, are still very much embedded in their natal material and thus their jets are best observed in the sub-millimeter and near-infrared (NIR) regimes. For example, the HH~212 jet is shown in Figure 2 in a NIR line of molecular Hydrogen. In the Class II (T Tauri) stage, YSOs become optically visible and their jets can be traced back to the jet launch region in forbidden emission lines (FELs) like [OI]$\lambda$6300 or [SII]$\lambda$6731. High angular resolution observations of classical T Tauri (CTT) jets have provided observational constraints to jet launching models with measurements of jet collimation, rotation, and of the physical conditions in the jets being the most critical (Ray 2007b). Jet launching models also make predictions on the efficiency of the jets, i.e. the ratio of the mass outflow and accretion rates ($\dot{M}_{out}$ / $\dot{M}_{acc}$). Thus studies have also focussed on measuring this ratio. As well as the collimated jet launched by the star-disk system, the material entrained and swept up by the jet is also observed, for example in the form of CO molecular outflows (Bachiller 2009). 


Much progress has now been made in understanding protostellar outflow activity, especially with regard to low mass YSOs.  This is due not only to advances in technology and observing techniques but also thanks to the multi-disciplinary approach combining observations, experiments, theory, and numerical simulations which has been adopted (Tsinganos, Ray $\&$ Stute, 2009). In spite of this progress, many questions remain and there are many compelling reasons for studying protostellar jets and pursuing a proper understanding of them. Firstly, they are the likely mechanism by which angular momentum is removed from the accreting protostar, thus the action of jets could explain why disks accrete and why young stars rotate slowly. The removal of angular momentum via the jet could also affect how solids are moved around in the accretion disk and, thus, future planet formation (Combet $\&$ Ferreira 2008). Secondly, the X-ray and UV emission from the jet collimation region could possibly impact on the disk and, thus, planet formation processes (Glassgold et al. 2004). Thirdly, the action of protostellar outflows could be used to explain the star formation rate in molecular clouds which is considered to be low and the core to star efficiency of $\sim$ 30$\%$ (Padoan et al. 2014). Finally, the properties of a jet can often reveal information about the unresolved central engine of the star. For example, a wiggling or precessing jet could point to the driver being an unresolved binary (Whelan et al. 2010).

\begin{figure}
\centering
\includegraphics[width=80mm,height=28mm,trim= 0cm 0cm 0cm 0cm]{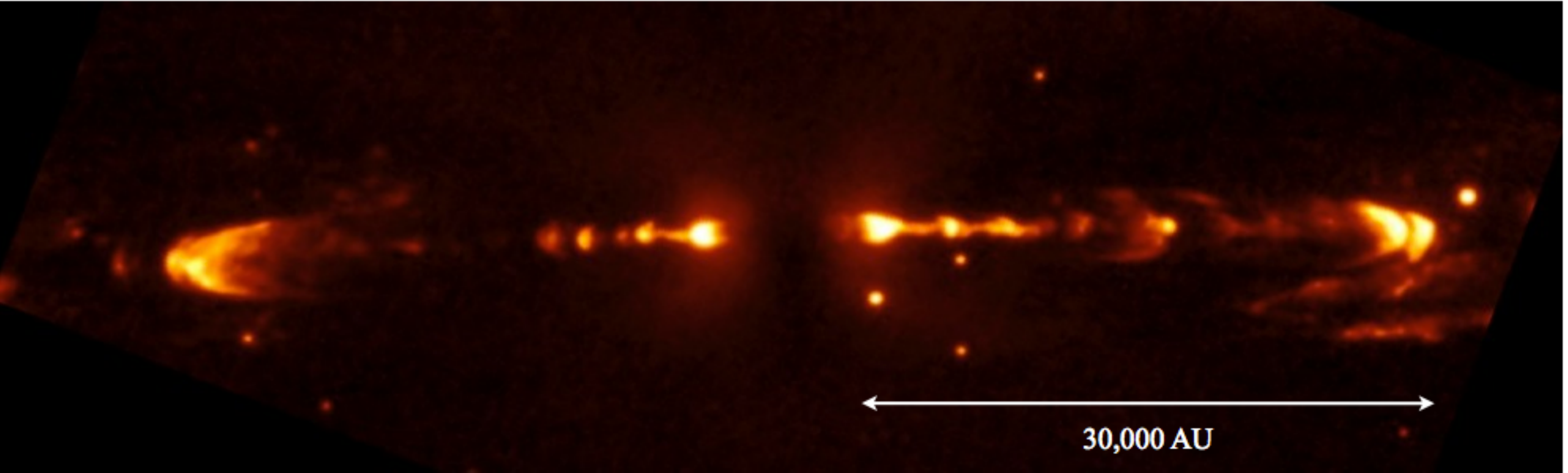}
\caption{The HH~212 jet in the H$_{2}$ 1-0 S(1) line at 2.122~$\mu$m taken from Zinnecker et al. 1998. Note that this jet is well collimated, made up of a series of shocks, several of which are bow-shock shaped and quite symmetric.}
\label{label2}
\end{figure}

The goal of this section was to provide a summary of what is currently known about protostellar jets and why they are important, with particular reference to low mass sources. Clearly this is a rich and vibrant field of study, a proper description of which would require much more space than is available here. Consequently, the rest of this article will be dedicated to a relatively new aspect of this field, jets from BDs. The detection of jets driven by young BDs is one of the most interesting developments in protostellar jet physics in the last 10 years. BDs are described as failed stars, due to the fact that they never accrete enough mass for Hydrogen burning. A non-negligible amount of the total mass of a star forming region will go into making BDs, thus it is important to understand how they form and evolve. The fact that young BDs have been found to have disks and outflows makes it likely that they form in a similar way to low mass stars and by investigating the properties of BD outflows information pertinent to BD formation and evolution can be provided. A further motivation for investigating BD outflow activity is to better understand outflow mechanisms in general. By probing outflow activity in objects with varying mass and $\dot{M}_{acc}$, better constraints can be placed on models describing jet launching and collimation. The mechanisms responsible for protostellar outflows are seen in a large range of astrophysical objects from BDs to active galactic nuclei (AGN) (a mass range of $\sim$ 10$^{8}$~\Msun). Hence a better insight into outflow mechanisms could have repercussions for several astrophysical disciplines from extragalactic astronomy to planetary science.

\section{Jets from Brown Dwarfs: Observations and Properties}

\subsection{Observing Brown Dwarf Jets}

It wasn't until the mid-1990's that telescopes became powerful enough to make the first observations of BDs. Now BDs are routinely observed, however, observations of their outflows can still be considered to be at the limit of what is achievable with state-of-the-art telescopes. As the rate of accretion in BDs is approximately two orders of magnitude lower than in CTTS, the luminosity of their outflows is expected to be fainter by the same factor. This estimate is supported by observations (Whelan et al. 2005). Due to their faintness it is necessary to observe BD outflows using 8~m class telescopes or better. The European Southern Observatory's (ESO) Very Large Telescope (VLT) has primarily been used for studies of BD outflows and an artists impression of a BD outflow published by ESO is shown in Figure 3. The first observations of BD outflows were spectroscopic in nature and confirmation of these observations relied on the technique of spectro-astrometry (SA). 
Since this early work, optical and molecular outflows from BDs have been imaged and the work of constraining $\dot{M}_{out}$ / $\dot{M}_{acc}$ for BD jets has also begun. Below, the results of the spectro-astrometric and imaging studies of BDs are summarised and the progress in understanding the efficiency of BD outflows discussed.

\begin{figure}
\centering
\includegraphics[width=55mm,trim= 2cm 0cm 2cm 1.5cm]{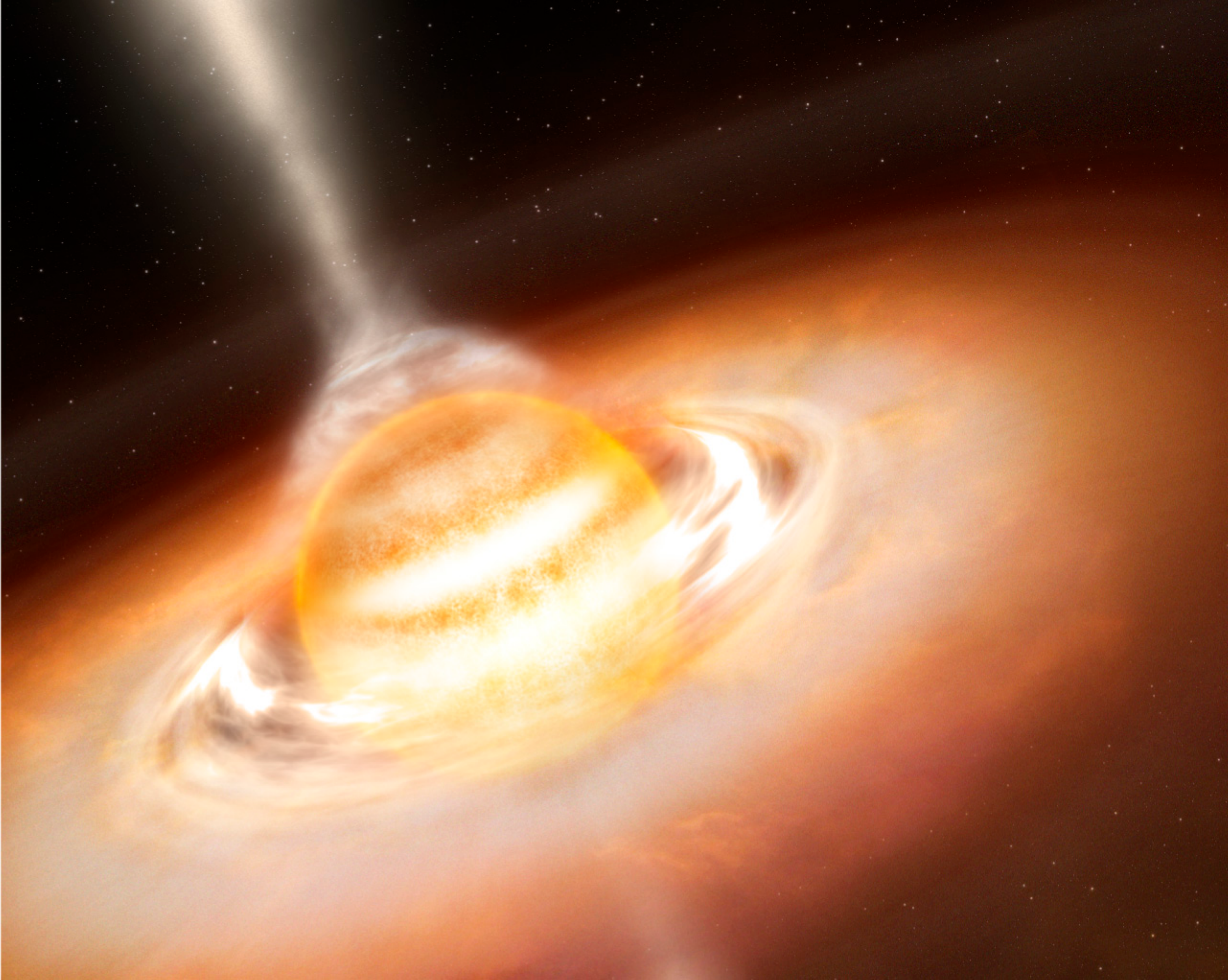}
\caption{Artist's Impression of a BD jet taken from the ESO press release http://www.eso.org/public/news/eso0724/ . This press release described the work of Whelan et al. 2007 which reported the detection of the 2MASS1207A jet. }
\label{label3}
\end{figure}


\begin{figure}
\centering
\includegraphics[width=82mm]{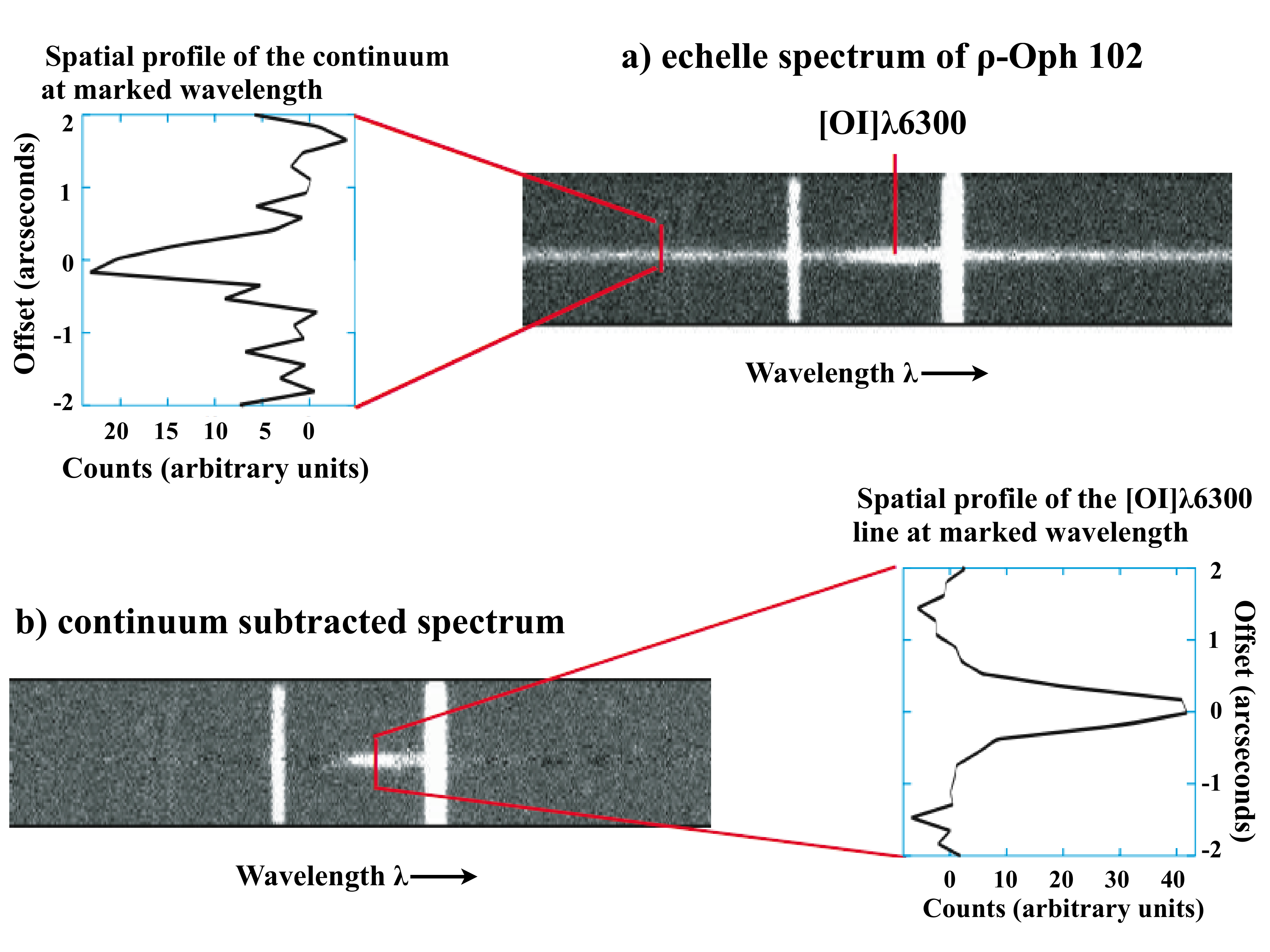}
\caption{Example of applying spectro-astrometry to a FEL spectrum of a BD. This Figure is adapted from Whelan et al. 2005. The region of the [OI]$\lambda$6300 line in the UVES spectrum of ISO-Oph~102 is shown here. The vertical lines are the sky lines. First the position of the continuum emission is measured as a function of wavelength. Typically the centroid of the spatial profile of the emission region is measured using Gaussian fitting. The second step is the subtraction of the continuum emission. The position of the pure [OI]$\lambda$6300 emission line can then be measured and plottied relative to the continuum position. The accuracy to which the centroid can be measured is dependent on the number of detected photons and is given by the formula $\sigma$ = $\frac{FWHM}{2.355\sqrt{N_{p}}}$ where N$_{p}$ is the number of detected photons.  }
\label{label4}
\end{figure}

\subsubsection{Spectro-astrometry}Spectroscopic studies of CTTSs have clearly demonstrated that optical and NIR FELs are important probes of outflows. These lines play a vital role in the cooling of the shocks, which result from the interaction of jets with their ambient medium. Hence, jets are frequently observed through their FEL regions. The first indication that BDs drive outflows came from the detection of FELs in their spectra (Fern{\'a}ndez $\&$ Comer{\'o}n 2001). However, the FEL regions were faint and not extended away from the BD to distances greater than the seeing. Typically, for CTTSs the FEL regions are extended up to tens of arcseconds from the star. Therefore, early spectroscopic observations of BDs could not confirm that the BD FEL regions originated in outflows. The approach adopted to solve this problem was to analyse the FEL regions using the powerful technique of SA (Whelan $\&$ Garcia 2008).  The UV-Visual Echelle Spectrometer (UVES) on the VLT was used and BDs which were known to be strong accretors were targeted. SA involves using Gaussian fitting to measure the centroid of the point spread function (PSF) of a spectrum as a function of wavelength, and thus recovers the relative position of  line and continuum emission to an accuracy dependent on the signal to noise. Thus spatial information below the seeing limit of the observation can be extracted (see Figure \ref{label4}). In other words, if a spectrum of a BD is obtained under seeing conditions of 1~\arcsec\ and an emission line region in that spectrum originates in outflow and is offset by 0".5 from the BD, this offset would not be detected unless the centroid of the spectrum is analysed using SA.

\begin{figure}
\centering
\includegraphics[width=70mm, trim= 2cm 0cm 2cm 1.5cm]{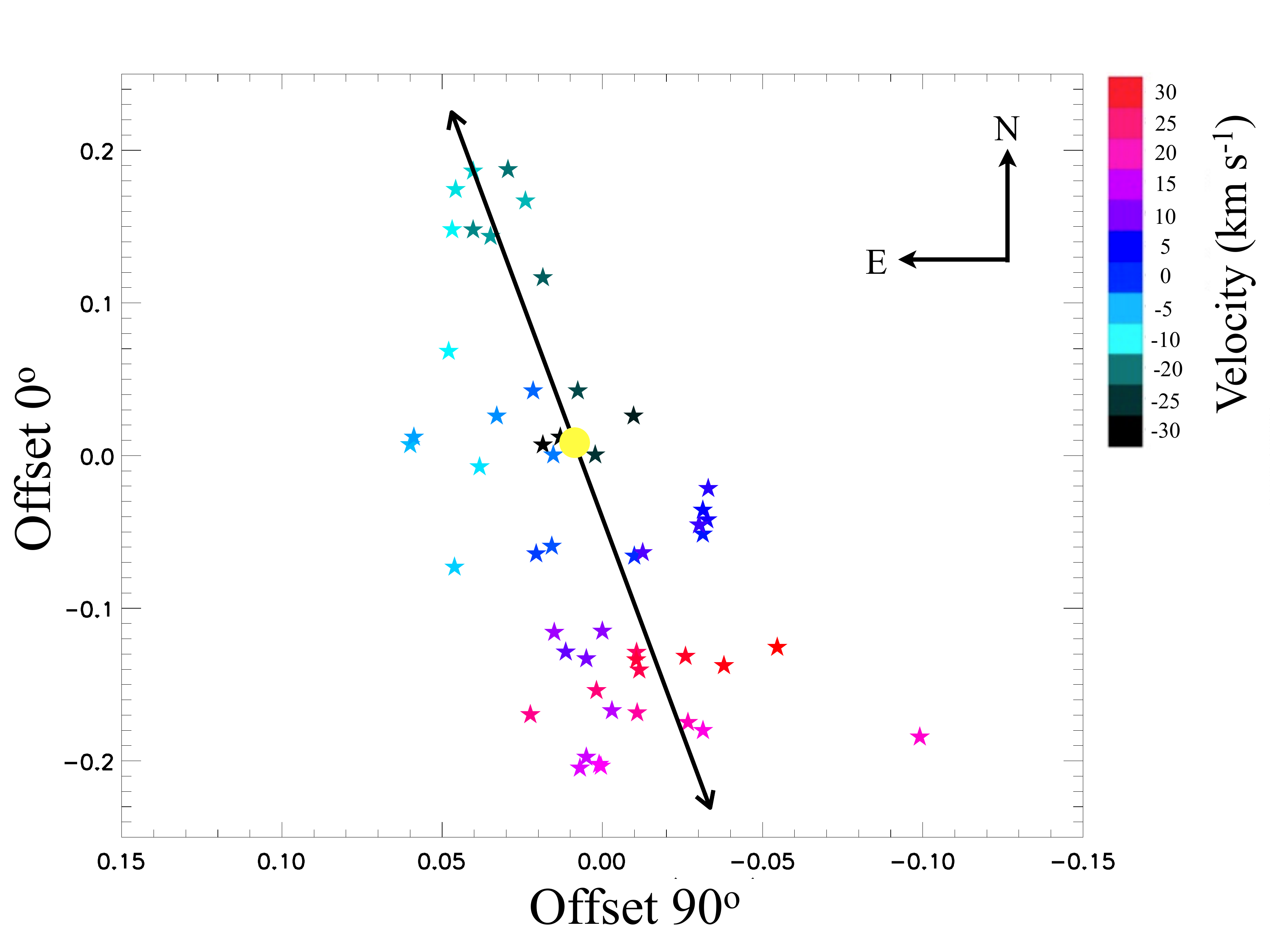}
\caption{Spectro-astrometric analysis of the ISO-Cha1 217 FEL regions. This figure is adapted from results presented in Whelan et al. 2009b. The blue-shifted outflow has a PA of $\sim$ 20$^{\circ}$. Also note that velocities of $\sim$ 30~km~s$^{-1}$ are reached in the red lobe while velocities in the blue lobe are approximately half that number.  Whelan et al. 2009b found this jet to be asymmetric in velocity.}
\label{label5}
\end{figure}

\begin{figure}
\centering
\includegraphics[width=105mm, angle=-90,trim= 0cm 0cm 0cm 0cm]{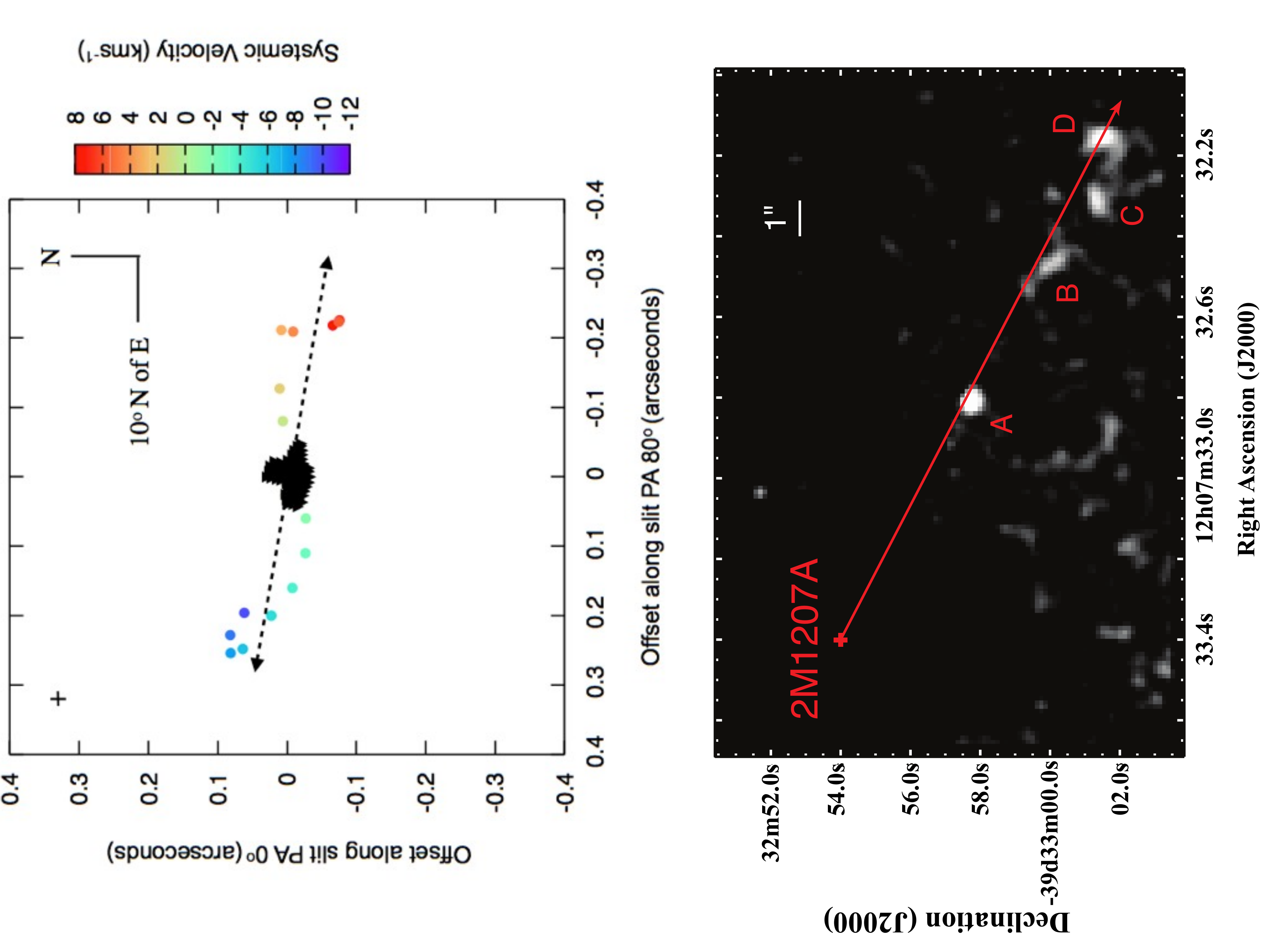}
\caption{Observations of the jet from the 24~\Mjup\ BD 2MASS1207A. Top Panel: Results of the spectro-astrometric analysis. The velocity of the outflow is $\sim$ $\pm$ 10~km~s$^{-1}$ and it has a PA of $\sim$ 65$^{\circ}$. Bottom Panel: [SII]-I image of the vicinity of 2MASS1207A. Several HH~objects are detected along the outflow PA with knot D having a bow shape with the apex pointing away from 2MASS1207A. Whelan et al. 2012 concluded that these are shocks in the 2MASS1207A, meaning that it is in fact collimated and therefore a jet. Object A was shown to be a star or galaxy.}
\label{label6}
\end{figure}

Using SA it was possible to retrieve spatio-kinematic information from the BD FEL regions proving their origin in outflowing material and to date $\sim$ 10 BD outflows have been detected in this way. Furthermore, by obtaining two spectra taken with the slits at orthogonal position angles (PA), the PA of the outflow on the sky can be measured. The results of analysing orthogonal spectra of the BD ISO-ChaI 217 with SA are presented in Figure \ref{label5}. Detections to date also include the 24 M$_{JUP}$ object 2MASS1207A (Whelan et al. 2012). For 2MASS1207A, red and blue-shifted [OI]$\lambda$6300 emission was observed and SA showed the emission to be displaced in opposing directions from 
the continuum position, consistent with a bipolar outflow (Figure \ref{label6}, top panel). In addition to spatial information, information on the radial velocity of the BD outflows could also be taken from early spectra. Radial velocities were found to vary between 10~km~s$^{-1}$ and 50~km~s$^{-1}$ and are consistent with the idea that jet velocities decrease with mass. Proper motion studies which will allow the velocities of BD jets to be measured are yet to be completed.

\subsubsection{Imaging} In addition to the spectroscopic investigations, recent work has explored, using a combination of narrow and broad band filters, imaging as a tool for studying BD outflows. This has been done, for example, in the case of the 2MASS1207A outflow. [SII] narrowband images and I and R broadband images of 2MASS1207A were obtained using the FORS imager on the ESO / VLT (Whelan et al. 2012). The outflow was detected in the [SII] images firstly as an extension in the PSF of 2MASS1207A and secondly as a series of knots along the known PA of the outflow (Figure 5, bottom panel). The outflow PA had been estimated from earlier spectro-astrometric studies (Whelan et al. 2012). These images revealed that BD outflows could be collimated, i.e. jet-like and episodic, as in CTT jets. Furthermore, two BD molecular outflows have been imaged in the sub-millimeter lines of CO (Monin et al. 2012, Phan-Bao et al. 2008). It is assumed that these are analogous to the molecular outflows driven by low mass stars.


\subsubsection{Measuring $\dot{M}_{out}$ and $\dot{M}_{acc}$} Knowing the outflow PA is important for follow-up spectroscopic observations, once the initial detection is made, as it means that the slit can be placed along the jet axis allowing the morphology of the jet to be mapped. If a bipolar outflow is detected, asymmetries will also be evident from the line profiles shapes of the outflow tracers. Also the fluxes of various key lines can be used to estimate $\dot{M}_{acc}$ and $\dot{M}_{out}$.  For example the Br$\gamma$ line is known to originate in infalling material and its luminosity is used to estimate $\dot{M}_{acc}$, while the [SII]$\lambda$6731 line is an important outflow tracer and its luminosity is used to estimate $\dot{M}_{out}$. As mentioned above the ratio $\dot{M}_{out}$ / $\dot{M}_{acc}$ is a measurement of the efficiency of an outflow i.e. how much of the infalling material is ejected in the outflow, and has been well constrained in CTTSs (Hartigan et al. 1995). The high angular resolution observations of CTT jets which have been used to constrain jet launching models in low mass stars are not currently possible for BDs. However, measurements of $\dot{M}_{out}$ / $\dot{M}_{acc}$ can be made and used to investigate if low mass jet launching models can also be applied to BDs.

\begin{figure}
\centering
\includegraphics[width=45mm, trim= 2cm 1cm 4cm 3cm]{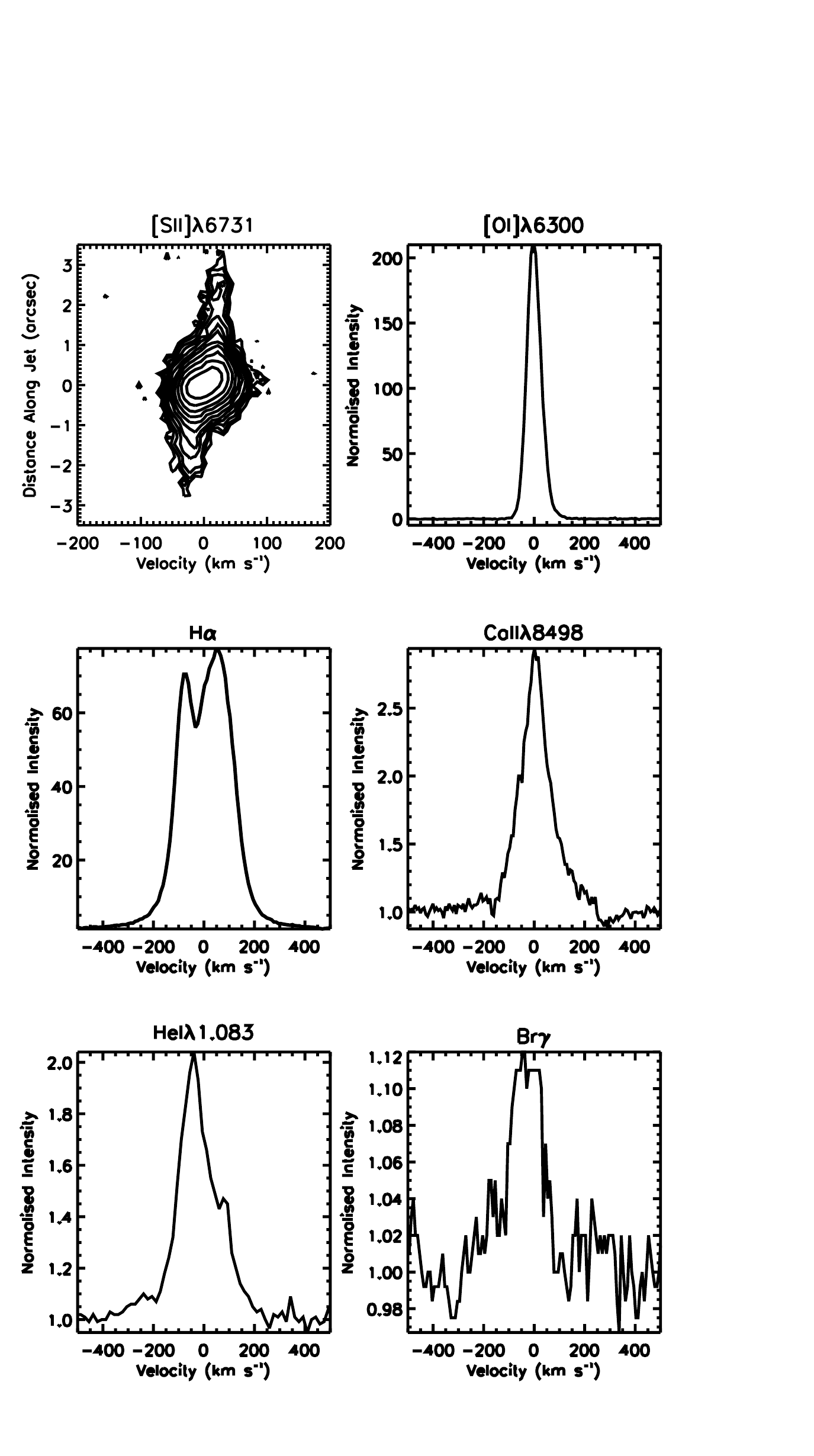}
\caption{Sample of the lines detected in an X-Shooter spectrum of Par-Lup3-4. These results are adapted from Whelan et al. 2014. In the top left panel a position velocity diagram of the Par-Lup3-4 jet is shown in the FEL of [SII]$\lambda$6731. Note the small scale of the jet ($\sim$ $\pm$3~\arcsec) and the low radial velocity of $\sim$ $\pm$ 20~km~s$^{-1}$. }
\label{label7}
\end{figure}

 The work of measuring this important ratio in BDs has begun and while the first results suggested that this ratio may be higher in BDs than CTTSs (Whelan et al. 2009b) recent new investigations have demonstrated that various biases can strongly influence these measurements (Whelan et al. 2014). This is well illustrated if we consider the case of the very low mass star (VLMS) Par-Lup3-4. In Figure 6 the jet of Par-Lup3-4 is shown in the [SII]$\lambda$6731 line, along with a sample of the outflow and accretion tracers detected in a recent X-Shooter spectrum of this object. X-Shooter is a multi-wavelength spectrograph on the ESO VLT. From these X-Shooter spectra Bacciotti et al. 2011 estimated $\dot{M}_{out}$ / $\dot{M}_{acc}$ at 30~$\%$ for Par-Lup3-4.  VLMSs have masses above the Hydrogen burning mass limit and below about $\sim$ 0.2~\Msun\ and, thus, this result supported the suggestion that $\dot{M}_{out}$ / $\dot{M}_{acc}$ increases with decreasing mass. A follow-on study of Par-Lup3-4 considered more carefully the effect of extinction on the measurements of both $\dot{M}_{out}$ and $\dot{M}_{acc}$. Additionally, the physical conditions in the jets were probed using various line ratios and the results combined with the luminosity of the [SII]$\lambda$6731 line to derive $\dot{M}_{out}$ through a calculation of the gas emissivity based on a 5-level atom model. This new work resulted in $\dot{M}_{out}$ / $\dot{M}_{acc}$ now being estimated at 0.048 (+0.097)(-0.019) for both the Par-Lup 3-4 red and blue jets.  These values agree with what has been measured for CTTSs and the example of Par-Lup3-4 provides a blue-print for future work on constraining $\dot{M}_{out}$ / $\dot{M}_{acc}$ for the lowest mass jet drivers. Projects are currently under-way to complete the goal of constraining $\dot{M}_{out}$ / $\dot{M}_{acc}$ in BDs. See Whelan et al. 2014 for a description of these results.

\subsection{Comparing Protostellar and BD Jets}
While the numbers of BD outflows detected to date is small and while they have not yet been observed in the same detail as CTT jets, it is still interesting to note the similarities between BD and CTT jets. Firstly, it is now apparent that BD outflows are well-collimated and it should be investigated if the collimation mechanism is the same as for low mass protostellar jets.  When the FEL regions of BDs were first shown to be extended in an outflow, the degree of collimation of these outflows was not known. Evidence of the collimation of BD outflows comes from the images of the 2MASS1207A jet shown in Figure 6 (Whelan et al. 2012) and from the detection of molecular outflows. It is theorised that molecular outflows are driven by collimated jets (Bachiller 1996) and thus their detection in the BD regime is indirect evidence that BD outflows can be jet-like. It is also clear from Figure 6 that BD jets can be episodic like protostellar outflows. Variable accretion has been studied in BDs (Scholz $\&$ Jayawardhana 2006) and it is likely that the different knots in the 2MASS1207A outflow represent different accretion events.

Looking again at the HH~212 jet shown in Figure 2, one of the most striking aspects of the jet is its symmetry. However, protostellar jets are rarely this symmetric and in fact are generally found to be asymmetric. Both kinematical and morphological asymmetries have been noted. Asymmetries have also been found in one of the BD jets studied to date, namely the jet from ISO-ChaI 217 (Whelan et al. 2009b, Joergens et al. 2012). Interestingly the red-shifted lobe of the ISO-Cha1 217 jet was found to be brighter and faster than the blue-shifted lobe. Normally the blue-shited lobe is the most prominent as the red-shifted lobe can be obscured by the accretion disk. The observations of the ISO-Cha1 217 jet point to ISO-ChaI 217 having a close to edge-on accretion disk. While the origin of jet asymmetries are not yet fully understood, these observations reveal that the mechanism which causes asymmetries in protostellar jets is also at work in the BD mass regime.  A further kinematical similarity between jets from BDs and low mass stars relates to the so-called low and high velocity components (LVC, HVC) observed in CTT jets. The FEL regions of CTTSs are known to have both LVCs and HVCs, with the HVC being the collimated jet and the LVC perhaps an uncollimated disk wind (Whelan et al. 2004). Evidence of the presence of HVCs and LVCs in the spectra of BDs comes from the study of LS-RCr~A1 (Whelan et al. 2009a).

One possible difference between the jets of the two mass regimes discussed here is with the efficiency of the outflows. As discussed in section 2.1.3, tentative evidence that $\dot{M}_{out}$ / $\dot{M}_{acc}$ is higher in BDs exists (Joergens et al. 2013b). If this is confirmed to be the case, it would be relevant to BD formation as it could be invoked to explain why BDs never accrete enough mass for H fusion to begin. However, this result will not be confirmed until the kind of investigation discussed in Whelan et al. 2014 has been completed for a significant number of young BDs with outflows. A second property unique to BD outflows is that the proper motion of the BD itself could be comparable to the proper motion of the outflow. A proper motion study is currently under-way for the 2MASS1207A outflow (Whelan et al. 2013). Preliminary results give a jet velocity of $\sim$ 20~km~s$^{-1}$ along a PA of 245$^{\circ}$. Song et al. 2006 report the proper motion of 2MASS1207A itself to be $\alpha$ = -62.7~mas and $\delta$ = -19.9~mas. The resultant motion is approximately 18~km~s$^{-1}$  along a PA of 252$^{\circ}$. Thus the proper motion of the jet is comparable to the proper motion of 2M1207A and is in the same general direction. An important consequence of this is that the driving source would have moved by a significant amount in between the ejections of different knots. This would affect their relative positions and thus the morphology of the jet. The example of 2MASS1207A shows that the proper motions of the BDs themselves could have an interesting effect on the low velocity jets they drive.

\section{Future Work}
As several important questions related to the role of jets in star and planet formation are still unanswered it is guaranteed that interest in the physics of jets will persist for some time to come. Future studies of protostellar jets will take advantage of advances in telescopes and observing techniques, e.g the Atacama Large Millimeter Array (ALMA) or the James Webb Space Telescope (JWST), and a multi-disciplinary approach to such studies will be key. For more information on protosteller jets and the direction of future studies, the chapter by Frank et al. from Protostars and Planets VI, ``Jets and outflows from star to cloud: Observations confront theory", is suggested reading. The lecture based on this chapter, delivered at PPVI by Sylvie Cabrit is also available to watch online \footnote{http://www.youtube.com/watch?v=5BJKGaBZrjw}. With regards to, future studies of BD jets work is very much limited at present by the faintness of the emission from the jets. In the immediate future the primary objective should be to concentrate on increasing the number of observed BD jets. By far the most efficient approach to this is to use imaging techniques to identify BDs with outflows. Follow-up spectroscopic studies can then be used to study the kinematics and physical conditions of the outflows. Once the sample has been increased studies aimed at constraining $\dot{M}_{out}$ / $\dot{M}_{acc}$ can then be seriously pursued. 


\acknowledgements{The author would like to acknowledge the input of her prinicipal collaborators, T.P. Ray, F. Bacciotti, J. Alcala and F. Comeron, to the results presented here.  E.T. Whelan also acknowledges support by the Research Grant Wh 172/1-1 of the Deutsche Forschungsgemeinschaft and the BMWi/DLR grant FKZ 50 OR 1309}

\end{document}